\documentclass[reprint,english,superscriptaddress,preprintnumbers,amsmath,amssymb,aps,prd]{revtex4-2}

\usepackage[utf8]{inputenc}

\usepackage{diagbox}

\usepackage{mathtools}
\usepackage{amsfonts}
\usepackage{mathrsfs}
\usepackage{bbm}
\usepackage{slashed}

\usepackage{graphicx}
\usepackage{color}
\usepackage{array}

\usepackage{booktabs}
\usepackage{makecell}
\usepackage{epstopdf}
\usepackage[caption=false]{subfig}

\usepackage{xspace}
\usepackage{siunitx}
\usepackage{hyperref}
\usepackage[nameinlink]{cleveref}

\usepackage{xifthen}
\usepackage{xcolor}
\hypersetup{
	colorlinks,
	linkcolor={blue!75!black},
	citecolor={blue!75!black},
	urlcolor={blue!75!black}
}

\setkeys{Gin}{width=0.48\textwidth}

\graphicspath{
	{./figures/}
	{../figures/}
}

\def\s0#1#2{\mbox{\small{$ \frac{#1}{#2} $}}}
\def\0#1#2{\frac{#1}{#2}}

\usepackage{multirow}
\usepackage{colortbl}
\definecolor{kugray5}{RGB}{224,224,224}
\newcommand{\PreserveBackslash}[1]{\let\temp=\\#1\let\\=\temp}
\newcolumntype{C}[1]{>{\PreserveBackslash\centering}p{#1}}
\newcolumntype{R}[1]{>{\PreserveBackslash\raggedleft}p{#1}}
\newcolumntype{L}[1]{>{\PreserveBackslash\raggedright}p{#1}}

\newcommand{\gettitle}{Soft modes in hot QCD matter}

\newcommand{\getHeidelbergAffiliation}{\affiliation{Institut f{\"u}r Theoretische Physik, Universit{\"a}t Heidelberg, Philosophenweg 16, 69120 Heidelberg, Germany}}

\newcommand{\getEMMIAffiliation}{\affiliation{ExtreMe Matter Institute EMMI, GSI, Planckstr. 1, D-64291 Darmstadt, Germany}}

\newcommand{\getJLUAffiliation}{\affiliation{Institute for Theoretical Physics, Justus Liebig University Giessen, Heinrich-Buff-Ring16, 35392 Giessen, Germany}}

\newcommand{\getHFHFAffiliation}{\affiliation{Helmholtz Research Academy Hesse for FAIR (HFHF), Campus Giessen, Giessen, Germany}}

\newcommand{\getHFHFAffiliationDA}{\affiliation{Helmholtz Research Academy Hesse for FAIR (HFHF), Campus Darmstadt, D-64289 Darmstadt, Germany}}

\newcommand{\getDarmstadtAffiliation}{\affiliation{Institut für Kernphysik, Technische Universität Darmstadt, D-64289 Darmstadt, Germany}}

\newcommand{\getDalianAFfiliation}{\affiliation{School of Physics, Dalian University of Technology, Dalian, 116024, P.R. China}}

\newcommand{\getUCASAFfiliation}{\affiliation{School of Nuclear Science and Technology, University of Chinese Academy of Sciences, Beijing, 100049, P.R. China}}
\newcommand{\getBITAFfiliation}{\affiliation{School of Physics, Beijing Institute of Technology, Beijing, 100811, P.R. China}}

\begin{document}

\title{\gettitle}

\author{Jens Braun}\getDarmstadtAffiliation\getEMMIAffiliation\getHFHFAffiliationDA

\author{Yong-rui Chen}\getDalianAFfiliation

\author{Wei-jie Fu}\getDalianAFfiliation

\author{Fei Gao}\getBITAFfiliation

\author{Chuang Huang}\getDalianAFfiliation

\author{Friederike Ihssen}\getHeidelbergAffiliation

\author{Jan M. Pawlowski}\getHeidelbergAffiliation\getEMMIAffiliation

\author{Fabian Rennecke}\getJLUAffiliation\getHFHFAffiliation

\author{Franz R. Sattler}\getHeidelbergAffiliation

\author{Yang-yang Tan}\getDalianAFfiliation

\author{Rui Wen}\getUCASAFfiliation

\author{Shi Yin}\getDalianAFfiliation


\begin{abstract}
The chiral crossover of QCD at finite temperature and vanishing baryon density turns into a second order phase transition if lighter than physical quark masses are considered. If this transition occurs sufficiently close to the physical point, its universal critical behaviour would largely control the physics of the QCD phase transition. We quantify the size of this region in QCD using 
the functional renormalisation group. 
This allows us to study both critical and non-critical effects on equal footing, facilitating a precise determination of the scaling regime. We find that the physical point is far away from the critical region. Importantly, we show that the physics of the chiral crossover is dominated by soft modes even far beyond the critical region. While scaling functions determine all thermodynamic properties of the system in the critical region, the order parameter potential is the relevant quantity away from it. We compute this potential in QCD using the functional renormalisation group and Dyson-Schwinger equations and provide a simple parametrisation for phenomenological applications.
\end{abstract}

\maketitle

\section{Introduction}
It has been argued that physical QCD at vanishing chemical potential could be in the critical scaling region of the chiral phase transition around  the limit of massless quarks \cite{HotQCD:2019xnw, Borsanyi:2020fev,Aarts:2020vyb, Kotov:2021rah, Aarts:2023vsf,Ding:2023oxy}. These studies have to be contrasted with analyses of QCD based on the functional renormalisation group (fRG) \cite{Fu:2019hdw, Braun:2020ada} that indicate a very small scaling region. The answer to this question is of phenomenological relevance as light low-momentum modes, so-called soft modes, dominate the dynamical evolution of the system created in a heavy-ion collision. If these modes exhibit critical scaling even at high beam energies, their contribution is well described by the respective universal critical dynamics, e.g.,~\cite{Hohenberg:1977ym, Berdnikov:1999ph, Rajagopal:1992qz, Grossi:2020ezz, Grossi:2021gqi}.

In the critical region, physical quantities such as the order parameter $\Delta$ are dominated by scaling behaviour, e.g., $\Delta(H) \sim H^{\frac{1}{\delta}}$, where $\delta$ is a universal critical exponent and $H$ an external field, see, e.g.,~\cite{Zinn-Justin:2002ecy}. For the chiral phase transition in QCD, $H$ can be identified with the light current quark mass. Crucially, this behaviour persists over many orders of magnitude in $H$. This poses the challenge both for theory and experiment that $\Delta(H)$ has to be known with high precision over a wide range of quark masses in order to make conclusive statements about universality. Furthermore, universality can only be used in the regime with negligible non-critical contributions. As demonstrated in \cite{Braun:2011sct}, conclusions about critical behaviour drawn from a scaling analysis may otherwise be incorrect. Therefore, it is clear that an analysis of the chiral critical behaviour with currently available lattice data \cite{HotQCD:2019xnw, Borsanyi:2020fev,Aarts:2020vyb, Kotov:2021rah, Aarts:2023vsf,Ding:2023oxy} requires great care.

In the present work we report on a conclusive scaling analysis with pion masses ranging over five orders of magnitude and as small as $m_{\pi}=0.002\,\text{MeV}$. This allows us to determine the size of the critical region of the chiral phase transition in QCD around the chiral limit. A key advantage of our renormalisation group based analysis is that fixed points are straightforwardly identified and the corresponding critical exponents can systematically be extracted with high precision within a given approximation \cite{Tetradis:1993ts, Balog:2019rrg, DePolsi:2020pjk, Dupuis:2020fhh}. Furthermore, universal behaviour only occurs in the linear regime around these fixed points \cite{Pelissetto:2000ek}, which uniquely defines the scaling region of the chiral phase transition of QCD.
This information is indispensable for a reliable application of universality to QCD, and hence is as relevant as the knowledge of universal properties.

The relevance of the magnetic equation of state $\Delta(H)$ and, more generally, the order parameter potential persists beyond the critical region. These quantities describe the universal and non-universal behaviour of the system and are the crucial input for the phenomenological modelling of heavy-ion collisions.
They are in general determined by the softest modes in the system.
In the present work, we provide the magnetic equation of state and the order parameter potential using functional approaches to QCD, i.e., the fRG approach and Dyson-Schwinger equations (DSE). We compute these quantities over many orders of magnitude in $H$ around and in particular below the physical point. The reliability of our results is corroborated by the excellent agreement of fRG and DSE results. We evaluate where soft modes dominate hot QCD matter and argue that their presence is the phenomenologically relevant property. Chiral criticality is only a very small aspect of this property and we will demonstrate that it is irrelevant for QCD at the physical point.

%
\begin{figure*}[t]
  \includegraphics[width=0.48\textwidth]{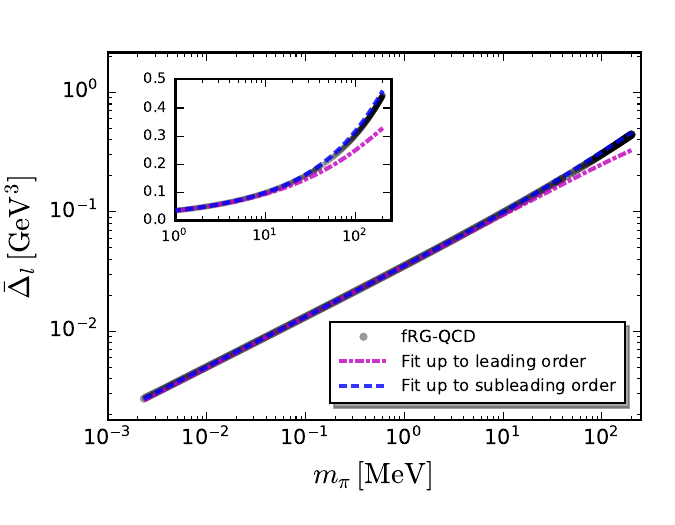}
  \hfill
  \includegraphics[width=0.48\textwidth]{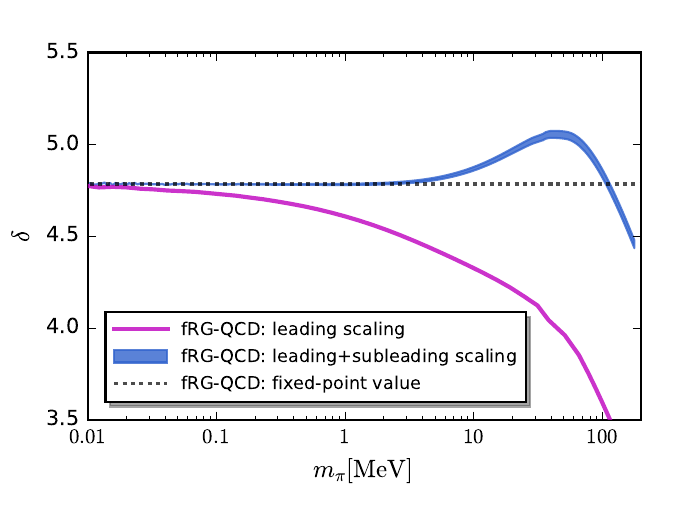}
\caption{\emph{Left panel:} Light chiral condensate as a function of the pion mass in comparison to fits based on \labelcref{eq:deltacrit}. \emph{Right panel:} Effective exponent $\delta$ as a function of the pion mass obtained from a scaling analysis of the fRG-QCD data in the left panel. The magenta line denotes the result extracted with only the leading scaling contribution to the chiral condensate in \labelcref{eq:deltacrit}, and the blue line includes the sub-leading correction in the pion mass. The blue band shows the error obtained from the uncertainty in determining the sub-leading scaling amplitude $a_m$. The dashed line indicates the critical exponent obtained from the fixed-point analysis of the fRG flows. For pion masses within the critical region the curves $\delta(m_\pi)$ must agree with this exponent, otherwise scale invariance is broken.
}\label{fig:Delta-mpi}
\end{figure*}
%

\section{Magnetic equation of state and the critical region}
The magnetic equation of state governs the $(T,m_q)$-dependence of the chiral condensate 
\begin{align}
\Delta_q(m_q\big) = m_q \frac{\partial \Omega\big(m_q\big)}{\partial m_q} = m_q \frac{T}{V}\int_x\!\big\langle \bar q(x)\,q(x) \big\rangle\,,
\label{eq:condensate}
\end{align}
associated with the quark flavour $q$, where \mbox{$\int_x = \int_0^{1/T}\!{\rm d}\tau\!\int\! {\rm d}^3x$} is the Euclidean spacetime integral and  $\Omega$  is the grand potential at finite temperature~$T$ over the spacetime volume $ V/T$.

In order to investigate the chiral properties of QCD, only the dependence on the light quark mass, $m_l$, matters. We therefore keep the strange quark mass fixed at its physical value. The relevant magnetic equation of state is then given by the light quark condensate $\Delta_{l}(m_l)$. Conventionally, the magnetic equation of state is defined with the universal scaling function $f_G(z)$ through the relation $\Delta_{l}/m_l = m_l^{1/\delta} f_G(z)$. Here, $z=(T-T_c)\, m_l^{-1/\beta\delta}/T_c$ is the scaling variable. 
The chiral properties of the magnetic equation of state are encoded in the magnetic susceptibility,
\begin{align}
\chi_M = -\frac{\partial \bar\Delta_l}{\partial m_l} \,, \qquad \textrm{with}\qquad \bar\Delta_l=\frac{\Delta_{l}}{m_l} \,.
\label{eq:chiM+barDelta}
\end{align}
We define the pseudocritical temperature through the peak location of $\chi_M(T)$. 

While the quark mass dependence of the pseudocritical temperature carries information about scaling, investigating the scaling behaviour of the chiral condensate is much cleaner since it is directly related to a universal scaling function. For a detailed discussion of the pseuocritical temperature we refer to \Cref{app:reg} and \ref{app:Tcdeterm}. Here, we compute the light chiral condensate $\bar\Delta_{l}(m_\pi)$ using the fRG for a range of pion masses spanning five orders of magnitude, $m_\pi \in [0.002,200]$\,MeV. This is shown in the left plot of \Cref{fig:Delta-mpi}. The fRG setup for QCD is explained in detail in \cite{Fu:2019hdw}. In order to study criticality we expand the magnetic equation of state at $T_c$ about $m_\pi = 0$, see \Cref{app:reg}. The chiral condensate in the critical region is then described by \cite{Zinn-Justin:2002ecy}
\begin{align}
\bar\Delta_{l}^{\rm (crit)}(m_\pi)= B_c\, m_\pi^{2/\delta}\big( 1+ a_m m_\pi^{2 \theta_H}
 \big)\,.
\label{eq:deltacrit}
\end{align}
We used that the relation between the current quark mass and the pion mass is $m_l \sim m_\pi^2$, in accordance with the Gell-Mann--Oakes--Renner relation. Here, we also included the sub-leading corrections to scaling, which is described by the universal exponent $\theta_H$. We emphasise that an accurate determination of the critical temperature $T_c$ is crucial for the validity of this expression \footnote{Otherwise, the leading correction due to deviations from $T_c$, which is $\sim m_\pi^{-2/\beta\delta}(T-T_c)$, needs to be taken into account. It is obtained by expanding the scaling form $\bar \Delta_{l}^{\rm (crit)} = m_\pi^{2/\delta} f_G(z)$ about $t=0$.}. This is detailed in \Cref{app:reg} and \ref{app:Tcdeterm}. If the system is in the critical region for sufficiently small $m_\pi \leq m_{\pi, \rm{crit}}$, the exponents $\delta$ and $\theta_H$ are universal, while the amplitudes $B_c$ and $a_m$ are not. In the critical region the chiral condensate is described accurately by \labelcref{eq:deltacrit} with a fixed set of fit parameters for all $m_\pi \leq m_{\pi, \rm{crit}}$. In fact, the size of the critical region can be defined unambiguously through this criterion.

A crucial advantage of the present RG setup is the fact that we can directly exploit critical scaling to accurately determine the critical exponents of the chiral phase transition. 
The set of RG flow equations of QCD in~\cite{Fu:2019hdw} can be reduced analytically to that of a 3$d$ $O(4)$ model in the scaling regime. 
We therefore find a second-order transition for vanishing light quark masses.
Note that this is in line with predictions from low-energy models, see, e.g., \cite{Pisarski:1983ms, Berges:1997eu, Schaefer:2004en, Fukushima:2008wg, Braun:2011sct, Resch:2017vjs}.
An analysis of the stability matrix of these RG equations at the Wilson-Fisher fixed point yields
\begin{align}\label{eq:ce}
\beta = 0.405\,,\quad
\delta = 4.784\,,\quad
\theta_H = 0.272\,,
\end{align}
for the magnetic and sub-leading exponents within the approximations used in this work. The computation of these exponents is detailed in \Cref{app:scale}. The small differences to the exact results are related to corrections of higher orders in a derivative expansion of the mesonic RG flows \cite{DePolsi:2020pjk}. They have been neglected here, and their small size supports the quantitative nature of the present analysis.

Within the critical region, \labelcref{eq:deltacrit} should accurately describe our data with the exponents in \labelcref{eq:ce} and fixed coefficients $B_c$ and $a_m$. In particular, at $T_c$ the critical exponent $\delta$ is  determined from
\begin{align}
\ln \frac{\bar \Delta_l(m_\pi)}{1+ a_m m_\pi^{2 \theta_H}} = \frac{1}{\delta} \ln m_\pi^2 + \ln B_c\,.
\label{eq:Deltatilde}
\end{align}
Note that the slope is independent of the leading amplitude $B_c$. 
We emphasise that this uniquely defines the scaling regime. Scale invariance at a critical point implies that thermodynamic functions are homogeneous. Using the RG, it can be shown that this arises in the regime where the RG flow equations can be linearised around the fixed-point solution. This procedure gives rise to all leading and sub-leading critical exponents, and a unique definition of the critical region. By using only the scaling form without regular corrections to analyse our data, we are precisely probing the extent of this linearised regime. This is, by definition, independent of the specific observable used and we choose the order parameter simply for convenience. We refer to the appendix for more details.

If only the leading scaling contributions are taken into account, we can use our results for $\bar \Delta_{l}(m_\pi)$ shown in the left plot of \Cref{fig:Delta-mpi} to extract the slope of \labelcref{eq:Deltatilde}. The resulting effective exponent $\delta(m_\pi)$ is shown as the magenta line in the right plot of \Cref{fig:Delta-mpi}. To guide the eye, the dashed line shows the value of the actual critical exponent $\delta$ of our system, see \labelcref{eq:ce}. In the critical region $\delta(m_\pi)$ must be independent of the mass and agree with this value. For the leading scaling behaviour this requires $m_{\pi} \lesssim 0.01$\,MeV.

The leading scaling behaviour can only be valid extremely close to the critical point. Within any reasonable numerical accuracy, sub-leading corrections can usually not be neglected. To take these corrections into account we need to determine the sub-leading amplitude $a_m$. To estimate the underlying uncertainty, we extract it from fitting our data of $\bar \Delta_{l}(m_\pi)$ with \labelcref{eq:deltacrit} and the exponents in \labelcref{eq:ce} in the range $m_\pi \in [0.01,m_{\pi,{\rm max}}]$ for two upper values $m_{\pi,{\rm max}} = 0.1$ and 10\,MeV. With the sub-leading exponent $\theta_H$ from \labelcref{eq:ce}, we can then use the full equation \labelcref{eq:Deltatilde} to extract $\delta(m_\pi)$ from our data. This is shown by the blue band in the right plot of \Cref{fig:Delta-mpi}. The boundaries of this band are obtained by using the sub-leading amplitudes $a_m$ extracted from the two different fit ranges defined above. We define the critical region as the range of pion masses where the resulting inverse slope $\delta(m_\pi)$ agrees with the actual exponent and find
\begin{align}
m_{\pi,{\rm crit}} \approx 2 - 5\, {\rm MeV}\,.
\end{align}
This result is rather insensitive to the precise value of $a_m$. In \Cref{app:fit} we present an error-based analysis of the critical region which yields a compatible result. 
In addition, in \Cref{app:fit} we show that with a combined analysis using critical and regular contributions, the accurate extraction of the critical $\delta$ is only feasible for $m_\pi \lesssim 25$\,MeV.
Hence, as a main result of this work, we observe strictly critical behaviour only for extremely light pions. The physical point of QCD is far away from $O(4)$-criticality.  This analysis also extends to the size of the critical region in temperature direction, which we estimate with $T-T_c \lesssim 1 - 7$\,MeV. We have deferred the respective analysis to \Cref{app:Tcdeterm} as it is beyond the main scope of the present work.

\section{Soft modes}
Our results clearly show that critical contributions are negligible in QCD near the physical point at vanishing chemical potential. Critical behaviour is driven by nearly massless critical modes. However, even without such critical modes, the presence of other soft modes, such as pseudo-Goldstone bosons, can significantly alter the dynamical behaviour of QCD, see, e.g., \cite{Son:1999pa, Grossi:2020ezz, Grossi:2021gqi}. Below, we shall show that soft modes drive the physics of the QCD phase transitions beyond the critical regime.

In the chiral limit, pions are exactly massless in the broken phase, since they are the Goldstone bosons of chiral symmetry breaking. In this case the phase transition is of second order and the pions and the sigma mode are massless at $T_c$. For non-vanishing light current quark masses, the pions acquire an explicit mass $m_\pi^2 \sim m_l$ in the broken phase. Still, for sufficiently small light quark masses they remain light in comparison to the other mass scales in QCD. In fact, they are the lightest particles in the low-energy spectrum of QCD \cite{ParticleDataGroup:2020ssz}. This facilitates the construction of systematic effective field theories such as chiral perturbation theory \cite{Gasser:1983yg}.

%
\begin{figure}[t]
  \includegraphics[width=1\columnwidth]{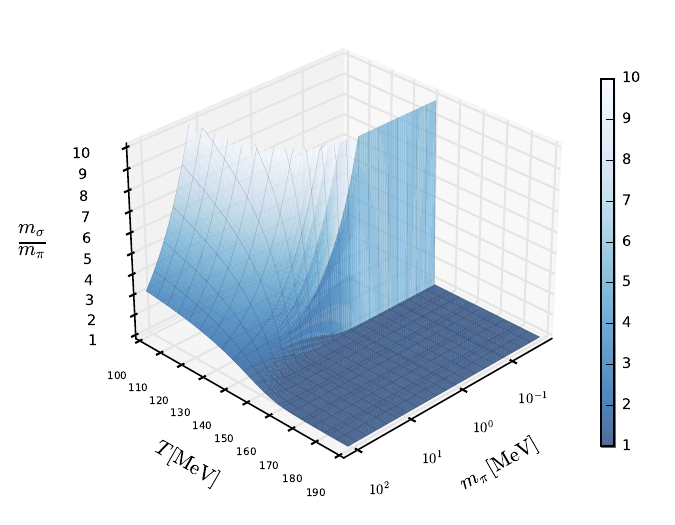}
\caption{Ratio of the $\sigma$ and $\pi$ meson masses $m_\sigma/m_\pi$ as a function of the temperature and the pion mass. The system has soft pion modes where this ratio is larger than one.}\label{fig:qmm}
\end{figure}
%

Hence, at low energies the structure of QCD is dominated by these soft modes. It is accurately described by pion exchange and the resulting effective theory is organised around the systematic expansion in powers of momenta $p^2$, which are, as a result of pion pole dominance, of the order of $m_\pi^2$. Such an expansion breaks down once $p^2$ or $m_\pi^2$ is of the order of the next-lowest resonances, which are the kaons and the $\sigma$ mesons \footnote{We tacitly identify the $\sigma$ mode with the $f_0(500)$ resonance, which is customary.}. If this is the case, exchange processes involving these particles become dominant. Since $1/m_\pi$ describes the correlation length of pions, the ratio $m_{\sigma/K}/m_\pi$ can be used as a measure of the softness of the pions: If this ratio is greater than one, chiral expansions are justified. 

These arguments are straightforward to verify in vacuum and for quantities that can be computed in Euclidean spacetime such as the equation of state in thermal equilibrium. The situation is more subtle for real-time and transport properties. In this case, the limits of vanishing frequency and the vanishing spatial momentum do not commute in a medium. A small-momentum expansion is therefore not guaranteed to be well-defined. Physically, this phenomenon is due to Landau damping, which gives rise to branch cuts whose location depends on the difference of particle energies \cite{Arnold:1992qy, Bellac:2011kqa}. For small mass differences, these cuts are present at small momenta. This not only means that simple small-momentum expansions in general cannot converge at finite $T$, but also that heavy particles can contribute to physical quantities through low-lying thresholds for  exchange processes with the heat bath. However, for every particle going on shell in such processes, there is always an off-shell contribution of at least one other particle. This naturally suppresses these contributions from heavy particles, as low-lying thresholds from heavy particles need to involve at least two heavy particles with a small mass difference. Relative to equivalent processes from light particles, the suppression is on the order of the ratio of light over heavy particle masses. Thus, although the systematic low-momentum expansion of chiral effective field theory is not always well-defined, low energy processes in QCD both in Euclidean and Minkowski space are still dominated by off-shell fluctuations of the lightest modes in the spectrum. 

%
\begin{figure}[t]
  \includegraphics[width=1\columnwidth]{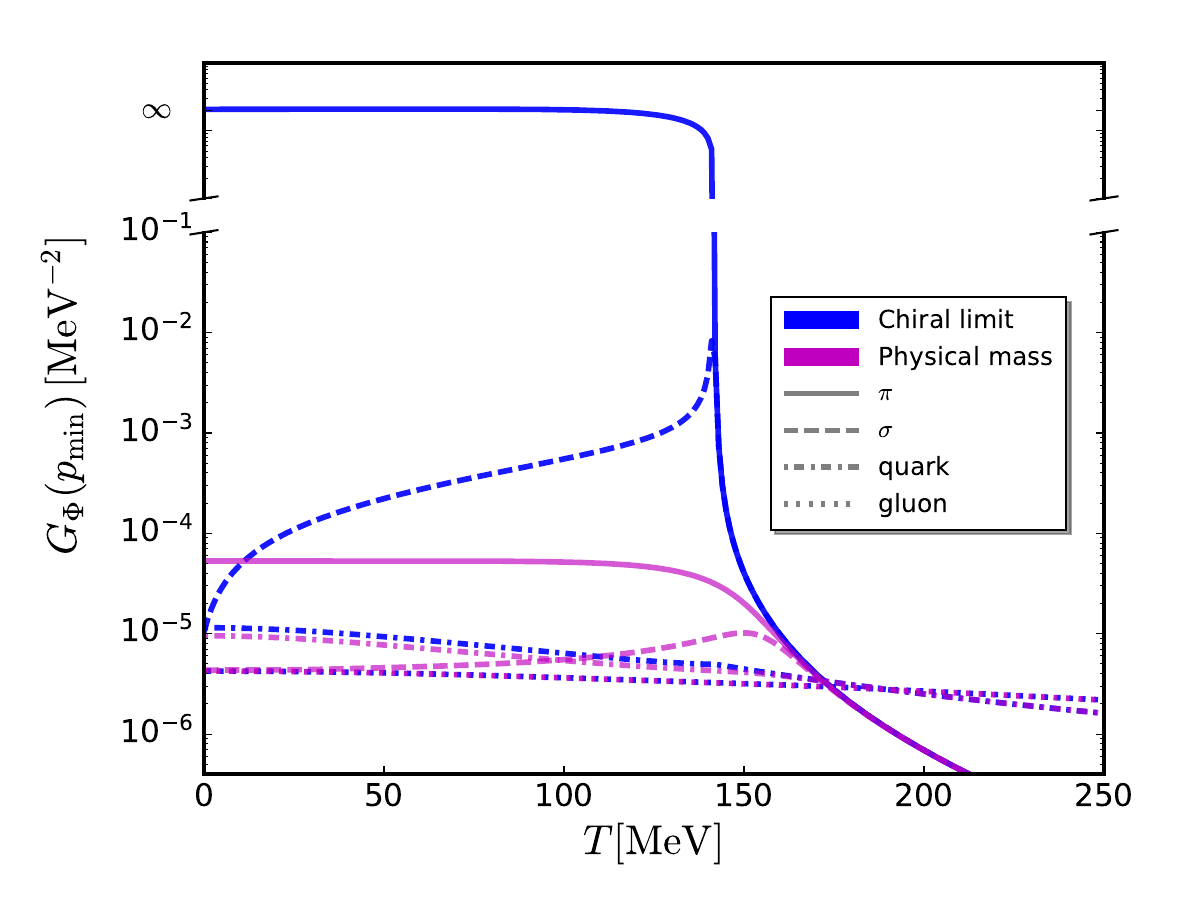}
\caption{Relevance of the contributions of quarks (dot-dashed), gluons (dotted) and pions (solid) and the $\sigma$ meson (dasehd) at finite $T$ in the chiral limit (blue) and at the physical point (pink). We estimate this relevance through the propagator $G_\Phi(p_{\rm min})$ of the field $\Phi = (\pi,\sigma,q,A)$, where $p_{\rm min}= (0,\boldsymbol{0})$ ($(\pi T,\boldsymbol{0})$) is the minimal Euclidean four-momentum of a boson (fermion) in a thermal bath.}\label{fig:gap}
\end{figure}
%

Naturally, the most striking example is provided by the critical region, which is exactly described by critical modes only. More importantly, a soft mode expansion, i.e., an expansion in terms of off-shell fluctuations of the lightest effective degrees of freedom, is far more general and is a powerful way to accurately describe the low-energy sector of QCD even beyond the critical regime.
	
The fRG setup used in the present work is geared towards describing the chiral phase structure of QCD, which is dominated by the light quarks. Consequently, $m_\sigma$ can be determined more accurately than $m_K$ in the present set-up. Hence, we use the ratio $m_\sigma/m_\pi$ as a measure for the validity of soft mode expansions. The result is shown in \Cref{fig:qmm} as a function of $T$ and the pion mass $m_\pi$, where the strange current quark mass is kept at its physical value and only $m_l$ is varied.

We see that $m_\sigma/m_\pi$ rapidly rises for temperatures below $T_{\rm pc}$.
In order to find a reasonable criterion to define the range of validity of soft mode expansions, we first consider the expectations for $m_\sigma/m_\pi$ in the critical region $H\in [0, H_{\rm crit}]$, where $\sigma$ is also a soft mode. In this region the ratio of the masses is universal and can be extracted from the numerical results in \cite{Engels:2003nq}. The largest value for this ratio at $z\geq0$ is reached at $T=T_c$, where one finds $\big(m_\sigma/m_\pi\big)(H)\big|_{H\leq H_{\rm crit}} \approx 2$, independent of $H$. It tends towards one with increasing $z$, e.g., increasing $T$ for fixed $H$. Thus at the pseudocritical temperature in the critical region one has 
\begin{align}
1<\left.\frac{m_\sigma}{m_\pi}(T_{\rm pc})\right|_{H\leq H_{\rm crit}}\lesssim 2\,.
\end{align}
Since the critical region is certainly well described by a soft-model expansion, we conclude that it should be valid whenever $m_\sigma/m_\pi > 1$ for $T\lesssim T_{pc}$\,.

To further illustrate the relevance of soft modes, we show the Euclidean propagator of different fields at minimal four-momentum in a thermal bath, $G_\Phi(p_{\rm min})$, in \Cref{fig:gap}. $G_\Phi(p_{\rm min})$ is an upper bound for the size of the propagator for real, Euclidean momenta \footnote{This is only true at vanishing density, since at finite density a moat regime can occur where nonzero spatial momenta are preferred \cite{Pisarski:2021qof, Fu:2024rto}.}.
It therefore is a direct measure for the size of the contribution of the respective field to the off-shell fluctuations in the system. The blue lines show the results in the chiral limit. We can clearly distinguish three regimes. For $T> T_c$ hadronic contributions are irrelevant and the dynamics of the system are dominated by quarks and gluons. At $T_c$ pions and the $\sigma$ become massless as a consequence of $O(4)$ critical behaviour. Thus, the immediate vicinity of $T_c$ is dominated by soft critical modes. Gluons and quarks are (thermally) gapped and hence play only a subleading role for $T \lesssim T_c$. Below $T_c$ the $\sigma$ is gapped by the nonzero chiral condensate, and we are left with the dynamics of massless pions as Goldstone bosons. This shows how the systems transitions from the critical regime to the regime accessible by chiral perturbation theory.

The pink lines in \Cref{fig:gap} show the relevance of different fields at the physical point. There are only small changes for quarks and gluons compared to the chiral limit. This is because the gluon contribution is dominated by the Yang-Mills mass gap and the quark contribution by the chiral condensate below $T_{\rm pc}$. As expected, the contributions of mesons are significantly suppressed for $T\lesssim T_{\rm pc}$. However, the qualitative picture remains the same as in the chiral limit: around the transitions pions and $\sigma$ are the dominant soft modes, while pions dominate at lower $T$. Note that the relevance of kaons is similar to $\sigma$ at lower temperatures, while around $T_{\rm pc}$, $\sigma$ mesons dominate over kaons as the former become lighter in this region, while the latter do not. The contributions of heavier hadrons will be suppressed even further, see, e.g., \cite{Rennecke:2015eba}. Thus, also away from the scaling regime we can clearly identify dominant soft modes.

\section{Order parameter potential}
We have established that hadronic matter around the physical point in QCD is largely controlled by soft modes rather than universality. Thus, while it is certainly appealing to rely on universality for phenomenological applications due to its simplicity, it is unwarranted. Instead of universal scaling functions, non-universal functions are required, and, in general, their computation is much more difficult in QCD. Of particularly broad interest is the order parameter potential $V_\chi(T,\bar \Delta_{l})$ as it fully characterises the static chiral properties of QCD. It contains all thermodynamic information as well as all correlations of the order parameter in terms of its derivatives. The latter cannot be extracted from the equation of state, but are indispensable input for dynamic modelling of heavy-ion collisions. For a first application in this direction, see \cite{Bluhm:2018qkf}.

Hence, due to its phenomenological relevance, we close this work by providing this potential.

It has been shown in \cite{Gao:2021vsf} how to compute the order parameter potential from the quark-mass dependence chiral condensate. Here, we briefly recapitulate the important steps: $V_\chi(\bar\Delta_l,T)$ is the Legendre transformation of the grand potential $\Omega(m_l,T)$ with respect to the light quark mass $m_l$,
\begin{align}
V_\chi(\bar{\Delta}_l,T)=\sup_{m_l}\Bigl[m_l \,\bar{\Delta}_{l}-\Omega(T,m_l)\Bigr]\,,  
\label{eq:LegV}
\end{align}
with the chiral condensate $\bar \Delta_l$ defined in \labelcref{eq:condensate,eq:chiM+barDelta}. Equation \labelcref{eq:LegV} implies $\partial V_{\chi}/\partial  \bar{\Delta}_{l}  = \,m_l$ and the $\bar\Delta_l$-integral of this relation provides us with $V_\chi$,
\begin{align}
 V_\chi(\bar{\Delta}_{l},T) = \int_{\bar{\Delta}_{l,\chi}(T)}^{\bar{\Delta}_l} d \bar{\Delta}\,  m_l(\bar{\Delta},T) \,.
\label{eq:VcompDelta}
\end{align}
The integral in \labelcref{eq:VcompDelta} requires the knowledge of $m_l(\bar{\Delta},T)$, which is obtained from $\bar{\Delta}_{l}(m_l,T)$ in \labelcref{eq:condensate} by inversion. In turn, for $\bar{\Delta}< \bar{\Delta}_{l,\chi}$ we have
\begin{align}
V_{\chi}(\bar{\Delta}_{l},T) =  V_{\chi}(\bar{\Delta}_{l,\chi},T)\,,
\end{align}
due to convexity of $V_\chi$ as a Legendre transform.

\begin{figure}[t]
	\includegraphics[width=0.5\textwidth]{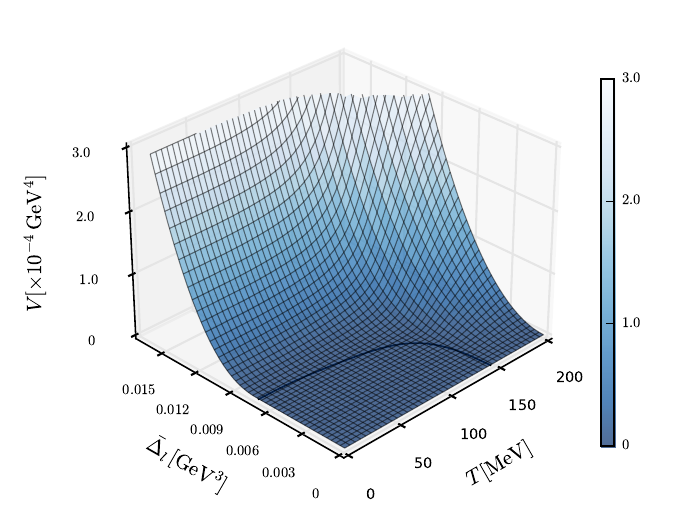}
	\vspace{-0.5cm}
	\caption{3D plot of the order parameter potential of QCD as a function of the light quark condensate and the temperature obtained with the fRG.}\label{fig:Poten-lightconden-3D}
\end{figure}

The numerical result for the effective potential from fRG-QCD is displayed in \Cref{fig:Poten-lightconden-3D}. That from the DSE computation is in quantitative agreement with the fRG result, which follows already from the quantitative agreement of the results for the renormalised condensate and its susceptibility from both methods, see, e.g., \cite{Gao:2021vsf}.

%
%

\section{Conclusions}
We have shown that the chiral crossover of QCD with physical quark masses is about two orders of magnitude in the pion mass away from the critical region of the chiral phase transition. We emphasise that a small scaling region facilitates accurate extrapolations based on non-critical data obtained, e.g., from heavy-ion experiments. However, around and below the pseudocritical temperature off-shell fluctuations of soft modes still dominate relevant physical aspects of the system. This is encoded in the order parameter potential, which we computed here from  functional approaches to QCD. This is the basis for \textit{beyond} Ginzburg-Landau type analyses of the chiral crossover regime in  QCD and a crucially important input for transport studies of the physics of Heavy-Ion collisions.

\begin{acknowledgments}
We thank B.~Brandt, H.~T.~Ding, C.S.~Fischer, O.~Philipsen and C.\ Schmidt for discussions. This work is done within the fQCD collaboration \cite{fQCD} and is funded by the National Natural Science Foundation of China under Grant No. 12175030, the Deutsche Forschungsgemeinschaft (DFG, German Research Foundation) under Germany’s Excellence Strategy EXC 2181/1 - 390900948 (the Heidelberg STRUCTURES Excellence Cluster) and the Collaborative Research Centre SFB 1225 - 273811115 (ISOQUANT), and the Collaborative Research Centers TransRegio CRC-TR 211 ``Strong-interaction matter under extreme conditions"-- project number 315477589 -- TRR 211. Moreover, JB acknowledges support by the DFG under grant BR 4005/6-1 (Heisenberg program).
\end{acknowledgments}



\appendix



\section{Scaling solution of QCD flows}
\label{app:scale}

Here we describe how we compute the critical exponents from the scaling solution of the fRG flow equations. The general strategy is to find the Wilson-Fisher fixed point associated to the second-order phase transition in the chiral limit. Scale invariance at this fixed point implies that the non-perturbative RG flow equations, which describe the RG-scale dependence of the (dimensionless) effective couplings, are zero. From this requirement, fixed point values for all couplings of the theory can be determined. By considering small deviations from these values, relevant (infrared-repulsive) and irrelevant (infrared-attractive) directions in the space of all running couplings can be identified and critical exponents can be extracted from their scaling with the RG scale $k$.

The full set of equations is given in the appendices of \ \cite{Fu:2019hdw}. Various simplifications can be made for the scaling analysis, as it has been shown in \cite{Mitter:2014wpa, Braun:2014ata, Rennecke:2015eba, Cyrol:2017ewj, Fu:2019hdw} that gluons and most hadrons decouple from the system in the infrared. To be specific, since we are considering a thermal phase transition, all fermionic degrees of freedom decouple and QCD reduces to a three-dimensional system of bosonic degrees of freedom. We emphasise that this not only entails gluons, but also bosonic resonances/bound states that are relevant in the vicinity of the phase transition. Formally, this is most easily seen in the Matsubara formalism of finite temperature field theory, where bosonic Euclidean frequencies are $ 2\pi n T/k$ and fermionic frequencies are $(2n+1)\pi T/k$ with $n \in \mathbb{Z}$. The scaling regime is reached in the deep infrared for $k\rightarrow 0$. Thus, at finite $T$ we have $T/k \rightarrow \infty$. Fermions do not have a Matsubara zero mode, but instead acquire a thermal mass $m_\psi/k \sim \pi T/k$ and hence decouple in the scaling regime. Similarly, all finite Matsubara modes of bosons decouple and only the zero mode with vanishing frequency survives. The resulting system only depends on spatial momenta and is therefore dimensionally reduced to $d=3$ dimensions. This is facilitated by the use of $3d$ spatial regulators here.

In addition, the gluons decouple in the infrared due to their finite mass gap \cite{Fu:2019hdw}. This leaves only mesons as potential relevant degrees of freedom. In the light chiral limit, all mesons involving strange quarks decouple as well, as they do not become critical at physical strange quark mass. This is hardwired in the present approximation, where the dynamics of these mesons do not feed back into the system of RG flows \cite{Fu:2019hdw}. The only mesonic interactions that are taken into account self-consistently stem from the scalar-pseudoscalar quark interaction channel, which gives rise to the pions and the $\sigma$ meson, i.e.,
\begin{align}
(\bar q q)^2+(\bar q\, i\gamma_5 \boldsymbol{\sigma} q)^2,
\label{eq:4qc}
\end{align}
with the Pauli-matrices $\boldsymbol{\sigma}$ in light-flavour space. The axial anomaly enters through the explicit breaking of $U_A(1)$ of this interaction channel. Thus, also the $\eta'$ meson is always massive in our case and decouples in the critical region. Lastly, we consider only cutoff scale $k$-dependent meson wave functions instead of taking into account the full momentum dependence. 

In summary, our system of QCD flow equations \emph{exactly} reduces to that of the 3$d$ $O(4)$ model in the LPA${}'$ truncation described below: The RG flow of the scale-dependent effective potential $U_k(\rho)$, with $\rho = \phi^2/2$ and the $O(4)$ vector field $\phi = (\sigma,\boldsymbol{\pi})$, in its scaling form is given by \cite{Berges:1995mw}
\begin{align}\label{eq:uflow}
\begin{split}
\partial_t u_k(\bar\rho) &= - d u_k(\bar\rho_k) + (d-2+\eta) \bar\rho_k u_k'(\bar\rho_k)\\
&\quad+ \frac{1}{2^{d-1} \pi^{d/2}\, d\,\Gamma\big(\frac{d}{2}\big)}\,\bigg(1-\frac{\eta}{d+2}\bigg)\\
&\quad\times\bigg(\frac{N-1}{1+u_k'(\bar\rho_k)}+\frac{1}{1+u_k'(\bar\rho_k)+2\bar\rho u_k''(\bar\rho_k)}\bigg)\,,
\end{split}
\end{align}
with $\partial_t = k\frac{d}{dk}$, $d=3$ and $N=4$. We defined $u_k(\bar\rho_k)= k^{-d} U\big( k^{d-2} Z_{\pi,k}^{-1}\, \bar\rho_k\big)$ and the dimensionless, renormalised field $\bar\rho_k = k^{2-d} Z_{\pi,k}\, \rho$, where $Z_{\pi,k}$ is the wave function of the pion. The anomalous dimension $\eta$ is given by
\begin{align}
\eta &= -\frac{\partial_t Z_{\pi,k}}{Z_{\pi,k}}= \frac{\bar\rho_{0,k} [u''(\bar\rho_{0,k})]^2}{2^{d-2} \pi^{d/2} \Gamma\big(\frac{d+2}{2}\big)}\,
\frac{1}{[1+2\bar\rho u_k''(\bar\rho_{0,k})]^2}\,.
\label{eq:etaeq}
\end{align}
In \labelcref{eq:etaeq}, $\bar\rho_{0,k}$ is the running minimum of the scale-dependent effective potential, i.e., the solution of the cutoff-scale dependent equation of motion
\begin{align}
u_k'(\bar\rho_{0,k}) = 0\,.
\end{align}
The critical effective potential $u^*$ follows from scale invariance at the fixed point,
\begin{align}
\partial_t u_k(\bar\rho)\big|_{u=u^*}=0\,.
\label{eq:fpeq}
\end{align}
\Cref{eq:uflow} is a partial differential equation which we reduce to a coupled set of ordinary equations by expanding the potential about its minimum,
\begin{align}
u(\bar\rho) = \sum_{n=2}^{N_u} \frac{\lambda_{n,k}}{n!}(\bar\rho-\bar\rho_{0,k})^n\,.
\label{eq:TaylorChiralLimit}
\end{align}
The ansatz \labelcref{eq:TaylorChiralLimit} is valid in the chiral limit. Away from it, we need to take into account a linear symmetry breaking source $ - c_\sigma \sigma$, with $c_\sigma$ being directly related to the current quark mass. We emphasise, however, that the second order transition in our case occurs in the chiral limit. This follows from the fact that we can exactly reduce our set of QCD flows to the 3$d$ $O(4)$ model in the scaling regime. The fixed point equation \labelcref{eq:fpeq} then reduces to a set of fixed point equations for the couplings $\lambda_{n,k}$ and the minimum $\bar\rho_{0,k}$. We use $N_u = 5$ throughout this work. Solving these equations using Newton's method yields the following fixed point values,
\begin{align}
\nonumber
\bar\rho_0^* &=0.07149\,, & 
\bar\lambda_2^* &=4.612\,, & 
\bar\lambda_3^* &=29.70\,, \\
\bar\lambda_4^*&=86.34\,, & 
\bar\lambda_5^*&=-1147\,. &
\end{align}
We emphasise that the fixed point values are not universal. But we can use them to extract the anomalous dimension by plugging them into \labelcref{eq:etaeq} and find
\begin{align}
\eta = 0.0373\,.
\label{eq:eta}
\end{align}
To extract further universal quantities, we perturb the couplings $\boldsymbol{g} = (\bar\rho_{0,k},\lambda_{2,k},\lambda_{3,k},\lambda_{4,k},\lambda_{5,k})$ around their fixed point values, $\boldsymbol{g} = \boldsymbol{g}^* + \delta\boldsymbol{g}$, and consider the linearised flow for small perturbations,
\begin{align}
\partial_t \delta\boldsymbol{g} \approx B^*\,\delta\boldsymbol{g}\,, \label{eq:linearisation}
\end{align}
where we used $\partial_t \boldsymbol{g}^* = 0$. $B^*$ is the stability matrix of the set of RG flows,
\begin{align}
B^*_{ij} = \frac{\partial (\partial_t g_i)}{\partial g_j}\bigg|_{\boldsymbol{g} = \boldsymbol{g}^*}\,.
\end{align}
Note that the flows $\partial_t g_i$ contain the anomalous dimension specified in \labelcref{eq:eta}.
The eigenvalues $b_i$ of the stability matrix determine the scaling of the couplings in the vicinity of the fixed point.
The eigendirection $\delta\bar g_i$ corresponding to the eigenvalue $b_i$ scales as

\begin{align}
\delta\bar g_i=\delta\bar g_{i,\Lambda}\left(\frac{m_l}{m_s},\frac{T}{T_c},\cdots\right)\,\cdot\, \left(\frac{k}{\Lambda}\right)^{b_i}\,,\label{eq:deltil-g}
\end{align}
where $\Lambda$ is some reference scale. One can see that all the couplings are homogeneous functions of the RG-scale $k$. The degree of homogeneity is given by the eigenvalues $b_i$, which are universal critical exponents. This implies scale invariance and is the origin of the scaling hypothesis \cite{Pelissetto:2000ek}. Note that the coefficients in \labelcref{eq:deltil-g} depend on external parameters, e.g., the temperature, quark mass and so on, and thus are not universal. Their magnitudes (relevant couplings, see below) determine whether the linearisation in \labelcref{eq:linearisation} is warranted, which in turn decides the linearised regime in the vicinity of the fixed point, that is, the size of the critical region.
For $b_i < 0$, the corresponding eigendirection is infrared-repulsive, while the directions with $b_i > 0$ are infrared-attractive. This defines relevant and irrelevant directions in the RG flow, respectively. The eigenvalues are in general complex, but we find three real ones,
\begin{align}
b_1 = -1.280\,,\qquad 
b_2 = 0.6745\,,\qquad 
b_3 = 2.722\,.
\label{eq:ev}
\end{align}
Hence, we find one relevant direction. Note, however, that we already used that this fixed point occurs in the chiral limit. Thus, there are actually two relevant directions, with the current quark mass being the second. To reach the second order transition, two parameters, the quark mass and, e.g., the temperature need to be fine-tuned. The relevant eigenvalue corresponds to the scaling of the screening mass of the critical mode and is therefore directly related to the correlation length exponent,
\begin{align}
\nu = -\frac{1}{b_1} = 0.781\,.
\label{eq:nu}
\end{align}
With two relevant universal exponents, $\eta$ and $\nu$, known, we can determine all other relevant critical exponents using scaling relations. For the present purposes, we need $\beta$ and $\delta$, which we get from
\begin{align}
\beta = (d-2+\eta)\frac{\nu}{2}\,,\qquad\qquad  \delta = \frac{d+2-\eta}{d-2+\eta}\,.
\label{eq:scalingrel}
\end{align}
The smallest irrelevant eigenvalue in \labelcref{eq:ev} gives the most important sub-leading critical exponent,
\begin{align}
\omega = b_2 = 0.6745\,.
\label{eq:omega}
\end{align}
and the sub-leading exponent used in the next-to-leading order (NLO) scaling analysis in main text is defined by
\begin{align}
\theta_H = \frac{\nu \omega}{\beta\delta} = 0.272\,.
\label{eq:thetaH}
\end{align}
Using the relations in \labelcref{eq:scalingrel,eq:thetaH} with the universal numbers in \labelcref{eq:eta,eq:nu,eq:omega} yields the exponents used in the main text in \labelcref{eq:ce}. With the second smallest irrelevant exponent in \labelcref{eq:ev} we can also define the exponent required for a next-to-next-to-leading order (NNLO) scaling analysis,
\begin{align}
\theta_3 = \frac{\nu b_3}{\beta\delta} = 1.097\,.
\end{align}
However, since $\theta_3/\theta_H \approx 4$, NNLO corrections are negligible.

The values we find are in excellent agreement with the most accurately determined exponents to date, see \cite{DePolsi:2020pjk} and references therein. We note that a systematic error analysis requires knowledge of higher-order corrections of the derivative expansion we used in meson sector of QCD. This is beyond the scope of this work. However, the good central values we find here are a testament to the rapid convergence of our truncation scheme.

\section{Critical and regular contributions to the equation of state}
\label{app:reg}

In general, we can express the relation between the chiral condensate and the magnetic equation of state as
\begin{align}
\Delta_l(T,m_\pi^2) = m_\pi^{2/\delta} f_G(z) + f_{\rm reg}(T,m_\pi^2)\,,
\label{eq:DeltaGen}
\end{align}
where first term is the singular contribution due to critical scaling and the second term is the regular contribution. While the critical contribution is a function of the scaling variable $z$, the regular part is an analytic function of both $T$ and $m_\pi^2$ independently. Note that we used that the external field is given by $H \sim m_l \sim m_\pi^2$. Expressing the critical part in terms of the leading and sub-leading scaling contributions leads to \labelcref{eq:deltacrit} if the regular part is neglected. We emphasise that universality can only occur if the latter is true. Since the regular part is analytic, it can be approximated as
\begin{align}
f_{\rm reg}(T,m_\pi^2) = b_r(T)\, m_\pi^2 +\mathcal{O}(m_\pi^4)\,.
\label{eq:fregExp}
\end{align}
We can also use \labelcref{eq:DeltaGen} to extract the contribution of the regular part to the pseudocritical temperature. To this end, we use that we define $T_{\rm pc}$ through the peak position of the temperature derivative of \labelcref{eq:DeltaGen},
\begin{align}
\frac{\partial \Delta_l(T,m_\pi^2) }{\partial T}= m_\pi^{2/\delta} f_G'(z) \frac{m_\pi^{-2/\beta\delta}}{T_{\rm c}} + \frac{\partial f_{\rm reg}(T,m_\pi^2)}{\partial T}\,.
\end{align}
Without regular corrections, this leads to a pseudocritical scaling variable $z_{\rm pc}$ with $f_G'(z_{\rm pc}) = 0$. A consistent definition of $T_{\rm pc}$ including regular corrections is then given by
\begin{align}
f_G'(z_{\rm pc}) + T_{\rm c} \frac{\partial f_{\rm reg}(T_{\rm pc},m_\pi^2)}{\partial T}\, m_{\pi}^{2/\beta\delta-2/\delta} = 0\,.
\end{align}
We see that the corrections from the regular contributions enter with an additional factor $m_{\pi}^{2/\beta\delta-2/\delta}$ and we find for the scaling of the pseuocritical temperature
\begin{align}
T_{\rm pc}(m_\pi^2) = T_{\rm c} + c\, m_\pi^{2/\beta\delta} + d_r(m_\pi^2)\, m_{\pi}^{2/\beta\delta-2/\delta}\,,
\end{align}
where $d_r(m_\pi^2)$ is an analytic function related to $f_{\rm reg}$. The form of the scaling contribution follows from the fact that in the critical region all universal quantities can be expressed as functions of a single scaling variable $z$. Similar to \labelcref{eq:fregExp}, expanding this function leads to
\begin{align}\label{eq:Tpcfull}
T_{\rm pc}(m_\pi^2) &= T_{\rm c} + c\, m_\pi^{2/\beta\delta} + c_r\, m_{\pi}^{2+2/\beta\delta-2/\delta} +\mathcal{O}(m_\pi^4)\,.
\end{align}
This agrees with Equation (8) in \cite{HotQCD:2019xnw}, and we can use this form for improving the fits for $T_{\rm pc}$. Note that there is no correction to $T_{\rm pc}$ from the sub-leading exponent $\theta_H$. This can be seen as follows. The chiral phase transition has two relevant parameters at criticality, which is reflected in the fact that we need to tune both the pion mass and the temperature in order to get to the second-order transition. The relevant parameters correspond to infrared-repulsive directions in the RG flow. The first two sub-leading (irrelevant) exponents correspond to the directions with the weakest infrared-attraction, i.e., the two smallest positive eigenvalues of the stability matrix of the set of RG flows, cf.\ \labelcref{eq:ev}. Taking only the relevant and these first two irrelevant directions into account, the free energy $f$ at criticality depends on three parameters $u_T$, $u_H$, $u_2$ and $u_3$ and behaves under rescaling with a factor $s$ as
\begin{align}\label{eq:scalingfunc}
f(u_T,u_H,u_2,u_3) = s^{-d} f\big(u_T s^{\frac{1}{\nu}},u_H s^{\frac{\beta\delta}{\nu}},u_2 s^{-\omega},u_3 s^{-b_3}\big)\,,
\end{align}
where we used the notation of \Cref{app:scale}.
The parameters $u_T$, $u_H$, $u_2$ and $u_3$ are linear combinations of effective couplings of the theory. The powers of $s$ on the right hand side are negative eigenvalues of the stability matrix, which we expressed in terms of known critical exponents. We can choose $s$ to absorb one of the arguments. For the magnetic equation we absorb $u_H$, i.e., $s=u_H^{-\frac{\nu}{\beta\delta}}$. Then the free energy scaling function becomes a function of only three arguments, $f_H(z,w,w_3)$, with
\begin{align}
z = u_T\, u_H^{-\frac{1}{\beta\delta}}\,, \qquad  w = u_2 u_H^{\theta_H}\,, \qquad  w_3 = u_3 u_H^{\theta_3}\,.
\end{align}
The parameters $u_T$ and $u_H$ receive their leading contributions from the reduced temperature $t$ and $m^2_\pi$. $w$ and $w_3$ are sub-leading scaling variables which depend on the irrelevant parameter $u_2$ and $u_3$ and the exponents $\theta_H$ and $\theta_3$ defined in \Cref{app:scale}. Since $\theta_H, \theta_3 >0$, the scaling variables $w$ and $w_3$ vanish for $m_\pi^2 \rightarrow 0$ and the magnetic scaling function only depends on the leading scaling variable $z$. Note that the dependence of $w$ and $w_3$ on $t$ is sub-leading \cite{Pelissetto:2000ek}, so we neglect these contributions to the magnetic scaling of $T_{\rm pc}$ in the main text.

From the scaling form of the free energy one can derive the magnetic scaling of the order parameter used in the main text. Including NNLO scaling effects it reads
\begin{align}\label{eq:fGNNLO}
\bar\Delta_l(m_\pi^2) = m_\pi^{2/\delta} f_G(z,w,w_3)\,.
\end{align} 
We emphasise that the global form of the leading-order scaling function $f_G(z)$ is well known, see, e.g., \cite{Karsch:2023pga} and references therein. Thus at $t=0$ only the leading scaling behavior $\bar\Delta_l(m_\pi^2) \sim m_\pi^{2/\delta}$ is encoded in $f_G(z)$. However, for a sensible scaling analysis in an extended region of pion masses at $T_c$, sub-leading scaling corrections need to be taken into account. By expanding \labelcref{eq:fGNNLO} to leading order in the subleading scaling variables and setting $t=0$ we find
\begin{align}
\begin{split}
\bar\Delta_l(m_\pi^2) &= B_c m_\pi^{2/\delta}\big(1+a_m m_\pi^{2\theta_H}+b_m m_\pi^{2\theta_3}\big)\\
&\quad+ \mathcal{O}\big(w^2,w_3^2\big)\,,
\end{split}
\end{align}
where the coefficients $B_c$, $a_m$ and $b_m$ are related to $f_G(z,w,w_3)$ and its derivatives at $t=0$.
As shown in \Cref{app:scale}, $\theta_3$ is much larger than $\theta_H$, so that the NNLO scaling correction $\sim m_\pi^{2\theta_3}$ is negligible. We verified explicitly that the determination of the size of the critical region presented in the main text is completely unaffected by this correction. We can therefore drop it and arrive at the scaling form given in Equation~(5) in the main text.

\section{Critical vs non-critical behaviour}
\label{app:fit}

For a precise determination of the critical region, we use that in the critical region for any $m_\pi \in [m_{\pi,{\rm min}},m_{\pi,{\rm max}}]$ with $m_{\pi,{\rm min}} < m_{\pi,{\rm max}} \leq m_{\pi,{\rm crit}}$, \labelcref{eq:deltacrit} should accurately describe our data with the exponents in \labelcref{eq:ce} and \emph{fixed} coefficients $B_c$ and $a_m$. In contrast, outside of the critical region a non-critical fit should describe our data more accurately. Typically, critical and non-critical contributions are taken into account in a single fit. In this case, the non analytic behaviour related to the phase transition should be stored completely in the critical part of the fit, involving in particular non-natural powers of $m_\pi$. All the regular contributions should be analytic and thus are polynomial in $m_\pi$. In order to be able to separate critical from non-critical behaviour in a clean way, we do not follow this strategy here. Instead, we use a fit that can only be valid outside the critical region. To this end, we use, as shown in \Cref{app:CompPotential}, that the order parameter potential $V(\Delta_{l})$ is well described by a low-order polynomial away from the critical point,
\begin{align}
V(\Delta_{l}) = \lambda_2 \Delta_{l}^2 + \lambda_4 \Delta_{l}^4+ \lambda_6 \Delta_{l}^6 - j \Delta_{l}\,.
\end{align}
The linear source term induces chiral symmetry breaking and is hence related to the pion mass, $j = c\, m_\pi^2$, with a dimensionful constant $c$ \cite{Fu:2019hdw, Gao:2021vsf}. The resulting equation of motion, $V'(\Delta_{l})=0$, thus fixes the relation between the pion mass and the chiral condensate,
\begin{align}
m_\pi^2 \sim 2\lambda_2 \Delta_{l} + 4\lambda_4 \Delta_{l}^3+ 6 \lambda_6 \Delta_{l}^5\,.
\end{align}
Taking the leading contributions of this relation into account leads us to the following ansatz for the non-critical order parameter,
\begin{align}
\Delta_{l}^{\rm (reg)}(m_\pi) = b_{\frac{1}{5}}\, m_\pi^{2/5}+ b_{\frac{3}{5}}\, m_\pi^{6/5} + b_{1}\, m_\pi^{2}\,.
\label{eq:deltareg}
\end{align}
We emphasise that this ansatz can only be valid outside the critical region for $m_\pi > m_{\pi,{\rm crit}}$. Despite the similarities of some of the exponents in \labelcref{eq:deltareg} to the critical ones in \labelcref{eq:ce}, e.g., $\delta \approx 5$ and $\theta_H \approx 0.3$, the branch cuts of $\Delta_{l}^{\rm (reg)}(m_\pi)$ at $m_\pi = 0$ are unrelated to the chiral phase transition since the origin lies outside of the range of validity of this function. In fact, it is the similarity of the exponents that makes the analysis of the critical region rather intricate: Without a large amount of high-precision data across a wide range of orders of magnitude, no conclusions about criticality can be drawn.
%
\begin{figure}[t]
  \includegraphics[width=1.0\columnwidth]{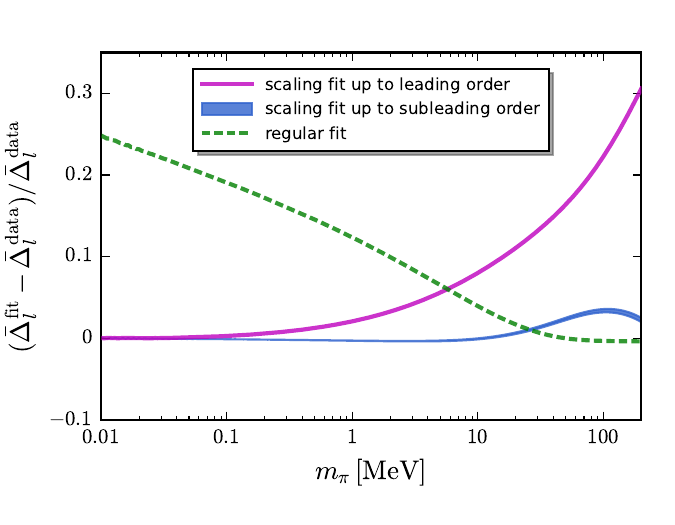}
\caption{Comparison of errors arising from the scaling and regular fits for the light chiral condensate.}
\label{fig:fiterror}
\end{figure}
%

In the critical region, the mass-dependence of the order parameter is better described by a scaling fit based on \labelcref{eq:deltacrit} with the exponents in \labelcref{eq:ce}, while outside the critical region a regular fit based on \labelcref{eq:deltareg} should yield higher precision. We use both scaling and regular fits for $\Delta_{l}(m_\pi)$. We employ two scaling fits in order to evaluate the relevance of the subleading scaling term: one fit only includes the leading term and the second one also includes the sub-leading term in \labelcref{eq:deltacrit}, similar as in the right panel of \Cref{fig:Delta-mpi}. The light quark chiral condensate $\Delta_{l}(m_\pi)$ in the range of $m_\pi \in [0.01,0.1]$\,MeV is used for the leading-order scaling fit, and those of $m_\pi \in [0.01,m_{\pi,{\rm max}}]$ for two upper values $m_{\pi,{\rm max}} = 0.1$ and 10\,MeV are used for the fit that also includes the sub-leading correction. The regular fits is done with the non-critical ansatz \labelcref{eq:deltareg} by using $\Delta_{l}(m_\pi)$ in the range of $m_\pi \in [100,200]$\,MeV. The resulting errors of the fits across the whole range of data are shown in \Cref{fig:fiterror}. We find that the error of regular fit is smaller than that of the leading-order scaling fit and that also including the subleading scaling for quark masses larger than $m_{\pi}\approx 6$\,MeV and $m_{\pi}\approx 25$\,MeV respectively.

We also emphasise that the quality of separate critical and regular fits as shown in \Cref{fig:fiterror} is only an indication for where scaling or non-scaling clearly dominates. However, there is a region where a combined fit including both scaling and regular contributions is required to most accurately describe the data. Together with our results of the main text, this region is approximately between 5 and 25 MeV. 
We confirmed this explicitly by a combined fit of the form
\begin{align}
\label{eq:deltacomb}
    \bar\Delta_l(m_\pi) = \bar\Delta_l^{(\rm crit)}(m_\pi) + r_1\, m_\pi^2 + r_2\, m_\pi^4\,,
\end{align}
where $\bar\Delta_l^{(\rm crit)}$ is given by \labelcref{eq:deltacrit} and the regular contribution in this case is modelled by a polynomial in the light quark mass \labelcref{eq:fregExp}. We performed this fit on data in the range $m_\pi\in [0.05,m_{\pi,\rm max}]$\,MeV. Indeed, for $r_1 = r_2 = 0$ the fit yields a critical exponent $\delta$ in agreement with the one in \labelcref{eq:ce} for $m_{\pi,\rm max} \lesssim 5$\,MeV. Inclusion of the regular contributions extends this range to $m_{\pi,\rm max} \approx 25$\,MeV. The precision of the data and the exponents is crucial here, as already an error of a few percent on $\delta$ will preclude conclusive statements about scaling, cf.\ \Cref{fig:Delta-mpi} and \ref{fig:fiterror}. Conversely, it is not possible to accurately extract $\delta$ from data with $m_{\pi,\rm max}\gtrsim 25$\,MeV with \labelcref{eq:deltacomb}.

In cases where very small pion masses are not available/accessible, a common strategy is to perform fits of the form \labelcref{eq:deltacomb} on data in mass ranges around the physical point. Our analysis shows that low order polynomials for the regular part as in \labelcref{eq:deltacomb} do not capture the mass dependence of the chiral condensate around the physical point accurately enough to allow for a meaningful extraction of $\delta$. The reason is that critical contributions are negligible close to the physical point so that the regular contributions cannot be described accurately by a low-order polynomial, cf. \Cref{eq:deltareg}. Thus, combined fits like \labelcref{eq:deltacomb} are not suitable for the extraction of universal quantities for data in ranges around $m_\pi \in [m_{\pi,min},140]$\,MeV for \emph{any} $m_{\pi,min}$.

Importantly, as soon as non-universal regular contributions are necessary to describe the data, the system itself is outside the universal scaling region. As shown in the main text, this occurs for $m_\pi \gtrsim 2-5$\,MeV. Our result in \Cref{fig:fiterror} is fully consistent with this finding. Furthermore, for $m_\pi \gtrsim 25$\,MeV, no information on scaling is required to accurately describe the data within the accuracy of our results. Importantly, this entails that even with the present high precision data no information about scaling can be extracted from data in this regime.

\section{Critical temperature}
\label{app:Tcdeterm}
 
In the critical region, the pseudo-critical temperature $T_{\rm pc}$ should behave as
\begin{align}\label{eq:Tpc}
T_{\rm pc}(m_\pi) \approx T_c + c\, m_\pi^p\,,
\end{align}
where $T_c$ is the critical temperature in the chiral limit, $m_l = 0$, and in the exponent $p$ is given by $p= 2/\beta\delta$, cf.\ \labelcref{eq:Tpcfull}. The mean-field values for these exponents are $\beta_{\rm MF} = 1/2$ and $\delta_{\rm MF} = 3$, and their exact values for the three-dimensional (3$d$) $O(4)$ universality class are $\beta = 0.387$, $\delta = 4.792$, and $\theta_H=0.307$, with the latter being the sub-leading magnetic scaling exponent, see \cite{DePolsi:2020pjk} and references therein. Interestingly, in the 3$d$ $O(4)$ critical region, $T_{\rm pc}$ would scale almost linearly with $m_\pi$ since $p = 1.079$.

 %
\begin{figure}[t]
  \includegraphics[width=1\columnwidth]{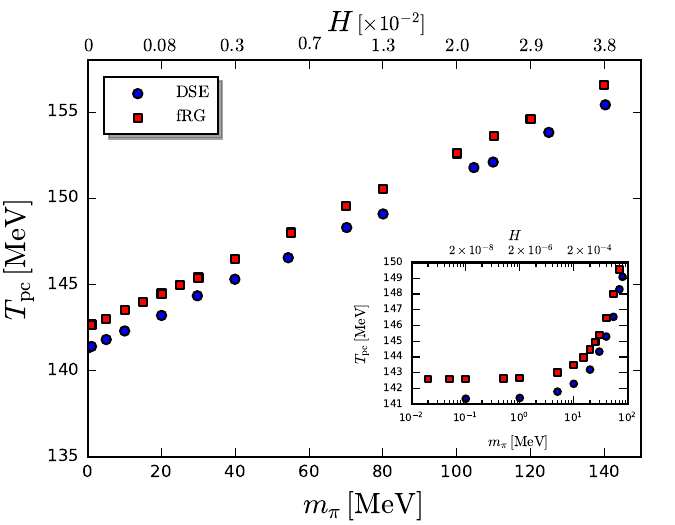}
\caption{Pseudo-critical temperature $T_{\rm pc}$ as a function of the pion mass obtained from functional QCD studies: this work (fRG) and \cite{Gao:2021vsf} (DSE). Here, $H=m_l/m_s$ with the strange current quark mass $m_s$ fixed at its physical value.}
\label{fig:Tc}
\end{figure}
%

In \Cref{fig:Tc} we show our results for $T_{\rm pc}(m_\pi)$. Both methods show excellent agreement. The fRG setup is explained in detail in \cite{Fu:2019hdw} and details on the DSE computation can be found in \cite{Gao:2021vsf}, see also \cite{Bernhardt:2023hpr} for a further very recent DSE study. By fitting these results according to \labelcref{eq:Tpc} across the whole range of available masses, one finds $T_{c,{\rm fRG}} = 142.568$\,MeV, $T_{c,{\rm DSE}} = 141.3$\,MeV, $c_{\rm fRG} = 0.08924$, $c_{\rm DSE} = 0.09581$, $p_{\rm fRG} = 1.024$, and $p_{\rm DSE} = 0.9606$. 
We note that the DSE setups used to date do not account for self-interactions of the critical mode which are however ultimately required to provide quantitative estimates of the size of the scaling region. A discussion of first important steps into this direction in DSE studies can be found below. In any case, without any further external input we therefore expect mean-field behaviour in the critical region with the DSE, i.e., $p_{\rm MF} = 4/3$.
This is an important observation regarding scaling analyses in general since it makes clear that the superficial agreement of results from different methods including the individual fit values for $c$ and $p$ and the effectively linear scaling of  $T_{\rm pc}(m_\pi)$ across many scales even beyond the physical point does not reveal critical behaviour at all. This finding supports the results of the main text.

\begin{figure}[t]
	\includegraphics[width=1.0\columnwidth]{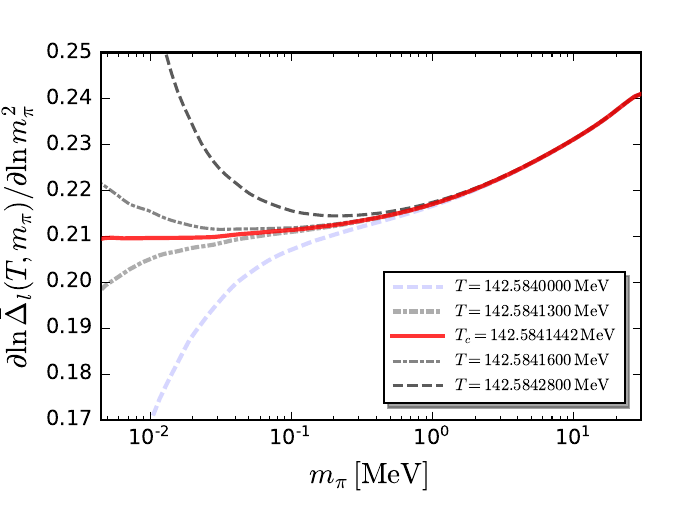}
	\caption{Determination of the critical temperature $T_c$ in the chiral limit through the behaviour of $\partial \ln \bar \Delta_l/\partial \ln m_\pi^2$ as a function of $m_\pi$ at several values of temperature in the vicinity of $T_c$.}\label{fig:Tc-determ}
\end{figure}

In \Cref{fig:Tc} we have used the pseudo-critical temperature $T_{\mathrm{ pc}}$ of the chiral crossover at finite quark mass to extract the critical temperature $T_c$ in the chiral limit by utilising \labelcref{eq:Tpc}. The resulting numerical precision of $T_c$, however, is not adequate to determine the critical behaviour of the phase transition, e.g., the critical exponents and the size of critical region. This is clearly demonstrated in \Cref{fig:Tc-determ}, where $\partial \ln \bar \Delta_l/\partial \ln m_\pi^2$ is shown as a function of the pion mass for several values of temperature in the proximity of the critical temperature. Obviously, if the temperature is $T=T_c$, the ratio should approach to $1/\delta$, as the red solid line shows. Once the temperature is away from the critical value, it deviates from the red curve quickly in the regime of small $m_\pi$. The larger $|T-T_c|$ is, the bigger is the $m_\pi$ where the deviation takes place. Consequently, one can use this property to determine the critical temperature to a very high precision, $T_c=142.5841442$ MeV, that is 10 significant numbers. 

%
%

\begin{figure}[t]
	\includegraphics[width=1.0\columnwidth]{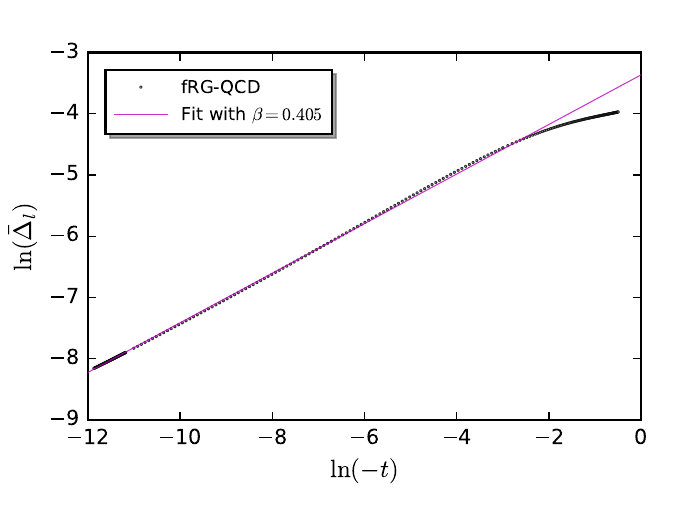}
	\caption{The logarithm of the light chiral condensate $\ln\Delta_l$ as a function of the logarithm of (minus) the reduced temperature $t=(T-T_c)/T_c$, see also \labelcref{eq:Delta-Tscaling}. }
	\label{fig:Delta-Tscaling}
\end{figure}

The present high precision data also allow us to estimate the size of the critical region in the temperature direction. A full analysis will be presented elsewhere. To this end, we study the dependence of the chiral condensate on the reduced temperature close to the chiral limit at $m_\pi = 0.002$\,MeV. To estimate the size of the critical region in temperature direction we perform a leading order scaling analysis using
\begin{align}
	\Delta_l(t) \approx A\, t^\beta\,,\quad \textrm{with}\;\; \beta\approx 0.405\,,\;\; t= \frac{T-T_c}{T_c}\,.
	\label{eq:Delta-Tscaling}
\end{align}
This form can be derived from the thermal scaling of the scaling function in \labelcref{eq:scalingfunc}. The exponent is determined in \Cref{app:scale}. In \Cref{fig:Delta-Tscaling} we show $\ln \bar\Delta_l$ as a function of the reduced temperature in comparison to $\beta \ln(-t)+\ln A$.  Strictly speaking the scaling terminates at about $\ln(-t) \approx -5$, but a sizable overlap still is visible up to  $\ln(-t) \approx -3$. This translates into a scaling region in the broken phase  with 
\begin{align}
	T_c-T \lesssim 1-7\,\textrm{MeV}\,,\qquad \textrm{for}\qquad T_c-T>0\,.
\end{align}
This is significantly smaller than the respective estimates from lattice QCD in \cite{Kotov:2021rah} and the very recent DSE study in \cite{Bernhardt:2023hpr}. The lattice study in \cite{Ding:2023oxy} finds a scaling region compatible with ours, but owing to a nonzero lattice spacing and the use of staggered fermions, an unphysical $3d$-$O(2)$ scaling is observed. We note that the leading scaling behaviour has also been used in these works. The effect of subleading corrections will be studied in a forthcoming work. 

%
%

In this context we emphasise once more that our present scaling analysis in functional QCD and the preceding one in \cite{Braun:2020ada} comes with two indispensable properties that are ultimately required to make conclusive predictions on the scaling and the size of the scaling regime:
\begin{itemize}
		\item[(1)] Scaling is generated dynamically.
		\item[(2)] Scaling properties of QCD are obtained for all masses with sufficiently high precision.
\end{itemize}	
While (1) is also present in lattice studies, the combination of (1) \& (2) can presently only be achieved within lattice simulations for large pion masses that are far outside the range of pion masses where we observe the dominance of critical scaling. Moreover, even with the present high precision data it is impossible to extract scaling properties for pion masses $m_\pi\gtrsim 25$\,MeV: while they have to be present as subleading corrections, their suppression is too large.

Other functional QCD studies often lack (1). This statement extends to the dynamics of the soft modes and their back reaction on the dynamics of the other fields. 
As mentioned in the main text, this holds true for all DSE studies to date, as self-interactions of critical modes have been neglected. However, first important steps in this direction have been taken in \cite{Fischer:2011pk, Gunkel:2021oya, Bernhardt:2023hpr}, where the feedback of critical and Goldstone modes is taken into account in the quark-gluon sector. For a detailed discussion of the required missing steps to fulfill (1) using DSEs we defer the reader to \cite{Gao:2021vsf}.

\section{Fit of the order parameter potential}
\label{app:CompPotential}

%
\begin{figure}[t]
	\includegraphics[width=1.0\columnwidth]{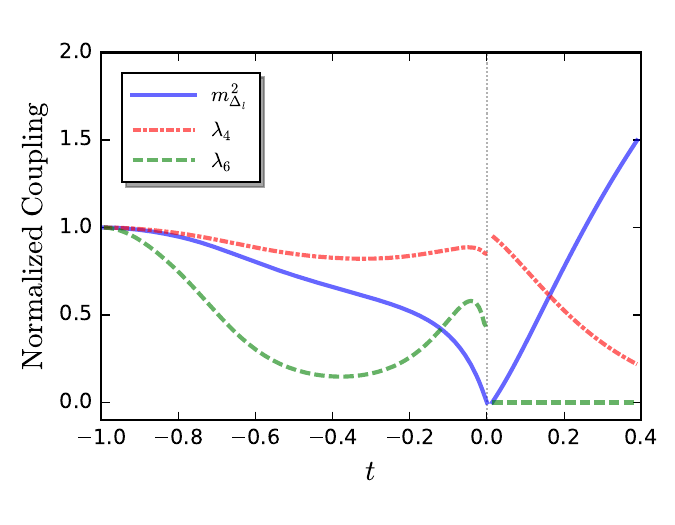} 
	\caption{Second, forth and sixth order couplings $m_{\Delta_{l}}^2(t)$ and $\lambda_{4,6}(t)$ as functions of the reduced temperature $t=(T-T_c)/T_c$, normalised to their respective values at $T=0$ for plotting. The are defined from separate fits to the effective potential in  \Cref{fig:Poten-lightconden-3D} in the chirally broken phase with \labelcref{eq:Vchi-broken}, and in the chirally symmetric phase with \labelcref{eq:Vchi-symm}. The couplings depicted here have a well-converging Taylor expansion in the reduced temperature defined in  \labelcref{eq:Coupling_t-expansions}, and the fit parameters are provided in \Cref{tab:poten-fit-coeffi}. The jump of the parameters at the phase transition originates in the separate fits in the broken and symmetric regime.}
	\label{fig:Poten-param}
\end{figure}
%

The current fRG study gives us direct access to the order parameter potential, whose scaling form in the chiral limit determines all critical properties. Note that this information is only indirectly obtained within lattice simulations, where one has to rely on the computation of correlation functions. They are defined as $m_q, \mu_B,...$-derivatives of the order parameter potential which comes along with a loss of accuracy and global information. We find that the dimensionless potential as a function of the light quark condensate in the broken phase can very accurately be parametrised as
\begin{align}
  \bar{V}_\chi(\tilde\Delta_{l})=&\frac{\lambda_4}{8}(\tilde{\Delta}_{l}^2-\tilde{\Delta}_{l,\chi}^2)^2+\frac{\lambda_6}{24}(\tilde{\Delta}_{l}^2-\tilde{\Delta}_{l,\chi}^2)^3\,,
  \label{eq:Vchi-broken}
\end{align}
with the dimensionless light quark condensate $\tilde{\Delta}_{l}=\bar\Delta_l/\bar\Delta_{l,\chi}(T=0)$, in the chiral limit $\tilde{\Delta}_{l,\chi}=\bar\Delta_{l,\chi}/\bar\Delta_{l,\chi}(T=0)$, where $\lambda_4\geq0$ and $\lambda_6\geq0$ are assumed. Here $\bar\Delta_{l,\chi}(T=0)=2.7 \times 10^{-3} \,\mathrm{GeV}^3$, and the potential in \labelcref{eq:Vchi-broken} is obtained from the dimensionful one by dividing a factor $\Delta_{l,\chi}(T=0) = 5.83 \times 10^{-8} \,\mathrm{GeV}^4$. Note that \labelcref{eq:Vchi-broken} is not convex. Convexity of the full potential as in \Cref{fig:Poten-lightconden-3D} can be implemented by multiplication with a theta function, $\bar{V}_\chi(\tilde\Delta_{l})\, \theta(\tilde\Delta_{l}-\tilde{\Delta}_{l,\chi})$.

The squared curvature mass at the minimum reads
\begin{align}
  m_{\Delta_{l}}^2=&\partial^2_{\tilde{\Delta}_{l}}\bar{V}_\chi(\tilde{\Delta}_{l}=\tilde{\Delta}_{l,\chi})=\lambda_4 \tilde{\Delta}_{l,\chi}^2\,.
\end{align}
In the symmetric phase the potential is well approximated by 
\begin{align}
  \bar{V}_\chi(\tilde\Delta_{l})=&\frac{m_{\Delta_{l}}^2}{2}\tilde{\Delta}_{l}^2+\frac{\lambda_4}{8}\tilde{\Delta}_{l}^4+\frac{\lambda_6}{24}\tilde{\Delta}_{l}^6\,.
  \label{eq:Vchi-symm}
\end{align}
These parametrisations are valid in the chiral limit. Finite light quark masses are included by adding a linear term $- m_l\, \bar\Delta_l$. At the physical point the current quark mass is $m_l=5$ MeV in our case.

%
\begin{table*}[t]
	\begin{center}
		\begin{tabular}{ccccccccccc}
			\hline\hline & & & & & & & & & \\[-2ex]
			&$n=0$& 1 & 2 & 3 & 4 & 5 & 6 &7 & 8 \\[1ex]
			\hline & & & & & & & & & \\[-2ex]
			$a$  &0  & -1.83 $\times 10^{4}$ &-7.03$\times 10^{4}$ & -1.38$\times 10^{5}$ &-1.23$\times 10^{5}$ &-4.06$\times 10^{4}$  &0 &0 &0 \\[1ex]
			$a'$ & 0 & 8.41$\times 10^{3}$ &3.27$\times 10^{4}$ & -5.63$\times 10^{4}$ &0 &0  &0 & 0 &0 \\[1ex]
			$b$  &2.88$\times 10^{3}$  &-9.54$\times 10^{2}$ & -1.32$\times 10^{4}$ &-3.78$\times 10^{4}$ &-3.79$\times 10^{4}$  &-1.29$\times 10^{4}$ &0  &0 &0 \\[1ex]
			$b'$  &3.27$\times 10^{3}$ &-7.23$\times 10^{3}$  &-5.71$\times 10^{3}$  &1.95$\times 10^{4}$   &0  &0  &0 &0 &0 \\[1ex]
			$c$  &1.39$\times 10^{5}$  &-2.24$\times 10^{6}$  &-3.69$\times 10^{7}$  & -2.14$\times 10^{8}$  & -6.52$\times 10^{8}$ & -1.14$\times 10^{9}$ &-1.14$\times 10^{9}$ & -6.06$\times 10^{8}$&-1.33$\times 10^{8}$ \\[1ex]
			$c'$  & 0&0 &0 &0 &0 &0 &0 &0 &0 \\[1ex]
			\hline\hline
		\end{tabular}
		\caption{Values of the expansion coefficients of an expansion in the reduced temperature $t$ of the second, forth and sixth order couplings $m_{\Delta_{l}}^2(t)$ and $\lambda_{4,6}(t)$ of the effective potential, defined in \labelcref{eq:Coupling_t-expansions}. The coefficients are computed from a fit to the numerical data presented in \Cref{fig:Poten-param}. }
		\label{tab:poten-fit-coeffi}
	\end{center}\vspace{-0.5cm}
\end{table*}
%

%
We use polynomials to fit the temperature dependence of coefficients in \labelcref{eq:Vchi-broken} and \labelcref{eq:Vchi-symm}  in the broken and symmetric phase, respectively. 
\begin{subequations}
\label{eq:Coupling_t-expansions} 
The temperature-dependent mass reads 
\begin{align}
\begin{split}
  m_{\Delta_{l}}^2(t)\Big|_{-t\in \{0, 1\}}&=\sum_{n=0}^{n_{max}}a_n (-t)^n\,,\\ 
  m_{\Delta_{l}}^2(t)\Big|_{t\in \{0, 0.39\}}&=\sum_{n=0}^{n_{max}}a'_n t^n\,,
  \label{eq:m2}
\end{split}
\end{align}
The mass is expanded in the reduced temperature $t=(T-T_c)/T_{c}$ with $t>0$ in the symmetric phase and $t<0$ in the broken phase.The fourth-order coefficients in the broken and symmetric phases are given by
\begin{align}
\begin{split}
  \lambda_4(t)\Big|_{-t\in \{0, 1\}}&=\sum_{n=0}^{n_{max}}b_n(-t)^n\,,\\ 
  \lambda_4(t)\Big|_{t\in \{0, 0.39\}}&=\sum_{n=0}^{n_{max}}b'_nt^n\,.
  \label{eq:m4}
\end{split}
\end{align}
Finally, rhe sixth-order coefficients in the broken and symmetric phases read
\begin{align}
\begin{split}
  \lambda_6(t)\Big|_{-t\in \{0, 1\}}&=\sum_{n=0}^{n_{max}}c_n(-t)^n\,,\\ 
  \lambda_6(t)\Big|_{t\in \{0, 0.39\}}&=\sum_{n=0}^{n_{max}}c'_nt^n\,.
  \label{eq:m6}
\end{split}
\end{align}
\end{subequations}
The numerical results are plotted in \Cref{fig:Poten-param}. They are used to determine the values of coefficients of the polynomials in \labelcref{eq:Coupling_t-expansions}, and the relevant results are collected in \Cref{tab:poten-fit-coeffi}. This can be used as input for phenomenological applications. We emphasise again that our finding that the order parameter potential away from criticality is well described by a sixth-order polynomial corroborates our choice for the fit to the non-critical order parameter in \Cref{app:fit}. We note that the fit polynomials in \labelcref{eq:Vchi-broken} and \labelcref{eq:Vchi-symm} cannot descibe the scaling regime, as the critical potential is a non-analytic function. The jump of the parameters in \Cref{app:fit} at the phase transition originates in the separate fits in the broken and symmetric regime and are also indicative of the  (exceedingly small) scaling regime surrounding it. Certainly, their phenomenological use is facilitated by smooth global fits. 
Respective polynomial fits and more elaborate ones, that also capture the small scaling regime, will be presented elsewhere.

\bibliography{ref-lib}

\end{document}